\documentclass[twocolumn,aps,prl,amsmath,showpacs]{revtex4}

\usepackage{graphicx}

\begin{document}

\title{Evidence for the weak steric hindrance scenario
in the \\
supercooled-state reorientational dynamics}
\author{Song-Ho Chong$^{1}$, 
Angel J. Moreno$^{2,3}$
Francesco Sciortino$^{4}$, 
and Walter Kob$^{5}$}
\affiliation{
$^{1}$Institute for Molecular Science,
      Okazaki 444-8585, Japan \\
$^{2}$Dipartimento di Fisica and INFM-CRS-SMC,
      Universit\`a di Roma ``La Sapienza'',
      P.le. A. Moro 2, 00185 Roma, Italy \\
$^{3}$Donostia International Physics Center, 
      Paseo Manuel de Lardizabal 4,
      20018 San Sebasti\'{a}n, Spain \\
$^{4}$Dipartimento di Fisica and INFM-CRS-SOFT, 
      Universit\`a di Roma ``La Sapienza'', 
      P.le. A. Moro 2, 00185 Roma, Italy \\
$^{5}$Laboratoire des Collo\"ides, Verres et Nanomat\'eriaux,
      Universit\'e Montpellier 2,
      34095 Montpellier, France}
\date{\today}

\begin{abstract}

We use molecular-dynamics computer simulations to study the translational
and reorientational dynamics of a glass-forming liquid of dumbbells. For
 sufficiently elongated molecules the standard strong steric
hindrance scenario for the rotational dynamics is found. However,
for small elongations we find a different scenario -- the weak steric
hindrance scenario -- caused by a  new type of glass transition in
which the orientational dynamics of the molecule's axis  undergoes a
dynamical transition with a continuous increase of the non-ergodicity
parameter. These results are in agreement with the theoretical predictions
by the mode-coupling theory for the glass transition.

\end{abstract}


\maketitle

Systems of particles interacting only via hard core potentials show,
in spite of their apparent simplicity, a variety of phases: fluids,
crystals, and glasses.  Several investigations have focused on the
transition between these phases upon a change of the packing fraction
$\varphi$.  Recent experimental and computational studies on systems
with {\em elongated} particles, like ellipsoids~\cite{Donev04} and
spherocylinders~\cite{Williams03}, have revealed interesting features
concerning their random (amorphous) jammed packing. E.g. it has been
found that for the latter systems the maximum attainable packing fraction
$\varphi_{\rm max}$ is a non-monotonous function of the elongation
parameter $\zeta$~\cite{comment-zeta-alpha} which characterizes the shape
of the particles~\cite{Donev04,Williams03}.  Starting, for $\zeta=0$,
at $\varphi_{\rm max} \approx 0.64$, the value for random close packing
of hard spheres, $\varphi_{\rm max}$ reaches a maximum $\gtrsim 0.70$ at
$\zeta \approx 0.5$ and then decreases again for larger $\zeta$.  For hard
core particles the quantity $\Delta \varphi= \varphi_{\rm max}-\varphi$
is the relevant parameter that governs the relaxation dynamics of the
system.  Hence it can be expected that the non-monotonous behavior of
$\varphi_{\rm max}$ has an important implication on the $\varphi$- and
$\zeta$-dependence of the dynamics, i.e. on the diffusion constant or the
$\alpha$-relaxation time.  For instance, it implies  that, if the shape of
the particle is changed at fixed $\varphi$, the dynamics can accelerate
or become slower. Since for small $\Delta\varphi$ the dynamics depends
very strongly on this difference, the relaxation times of the system
can change by orders of magnitude upon a change of $\zeta$ of just a few
percent, which thus can have important consequences for the mechanical
properties of the jammed packing or the crystallization dynamics.

From the theoretical side it is possible to understand the observed
non-monotonous behavior of $\varphi_{\rm max}(\zeta)$ at least
qualitatively within the framework of mode-coupling theory (MCT). For
the case of ellipsoids or dumbbells the theory predicts that the
critical packing fraction $\varphi_{\rm c}(\zeta)$, at which the system
undergoes a dynamical arrest, shows a maximum at intermediate values
of $\zeta$~\cite{Letz00,Chong-HDS}.  Although MCT predictions concern
the line $\varphi_{\rm c}(\zeta)$ and not $\varphi_{\rm max}(\zeta)$,
it can be expected that these two curves track each other closely and
that therefore the theory does indeed provide an explanation for the
experimental findings.

MCT also predicts that features of the {\em reorientational}
dynamics strongly depend on $\zeta$.  It is predicted, if $\zeta$
exceeds a certain threshold, $\zeta > \zeta_{\rm c}$ 
($\zeta_{\rm c} = 0.345$ for hard dumbbells~\cite{Chong-HDS}), 
that a dynamic arrest takes
place at $\varphi_{\rm c}(\zeta)$ which is usually denoted as ``type-B
transition''~\cite{Goetze91b}.  In supercooled states near $\varphi_{\rm
c}(\zeta)$, orientational correlation functions for angular-momentum index
$\ell$, $C_{\ell}^{\rm (s)}(t)= \langle P_{\ell}[{\vec e}(t) \cdot {\vec
e}(0)] \rangle$, show a two-step decay with a {\em finite} plateau height
(here $P_{\ell}$ is the Legendre polynomial of order $\ell$ and ${\vec
e}(t)$ is the orientation of a molecule at time $t$.)  When $\zeta$
is significantly larger than $\zeta_{c}$, the height of this plateau is
predicted to be larger, the relaxation time to be longer, and the final
$\alpha$-decay to be less stretched for $\ell=1$ than for $\ell=2$.
Such a scenario is called ``strong steric hindrance scenario'', and all these
features are in qualitative agreement with the experimental results
found by dielectric-loss ($\ell=1$) and depolarized-light-scattering
($\ell=2$) spectroscopy for propylene carbonate~\cite{PC-strong-all}
and glycerol~\cite{glycerol-strong-all}, and also with simulation results
for water~\cite{Fabbian98} and OTP~\cite{Rinaldi01}.

When particles are more spherical, i.e.  $\zeta < \zeta_{\rm c}$, MCT
predicts a novel different type of dynamical transition for odd-$\ell$
reorientational correlators~\cite{Letz00,Chong-HDS,Goetze00c}.
Translational and even-$\ell$ reorientational dynamics are predicted
to undergo a dynamical arrest at a packing fraction $\varphi_{\rm
c}(\zeta)$, whereas odd-$\ell$ correlators remain ergodic.  Only at
a higher packing fraction $\varphi_{\rm A}(\zeta) > \varphi_{\rm
c}(\zeta)$ the latter are predicted to undergo a dynamical arrest,
i.e. the orientation of the molecule's axis ($\ell = 1$) freezes
into a random direction.  Most remarkably it is predicted that this
second transition is of ``type-A'', i.e. the height of the plateau
in the odd-$\ell$ reorientational correlator is a smooth function of
$\varphi$, zero for $\varphi< \varphi_{\rm A}(\zeta)$ and increases
linearly for $\varphi> \varphi_{\rm A}(\zeta)$.  This is in contrast to
the translational and even-$\ell$ rotational correlators which exhibit
at $\varphi_{\rm c}(\zeta)$ a type-B transition, i.e. the height of
the plateau already takes a finite value at $\varphi_{\rm c}(\zeta)$.
The scenario for the reorientational dynamics of small elongations
which is caused by the nearby type-A singularity has been termed ``weak
steric hindrance scenario'' since, in the range $\varphi_{\rm c}(\zeta)
< \varphi < \varphi_{\rm A}(\zeta)$, the odd-$\ell$ correlators relax
even if the translational and even-$\ell$ reorientational dynamics are
frozen~\cite{Goetze00c}.  In this case, the $\alpha$-relaxation dynamics
is predicted to be strongly modified for the $\ell = 1$ reorientational
dynamics, whereas no such modification is predicted for $\ell = 2$.
So far the predicted existence of this type-A transition has not been
investigated neither by experiments nor computer simulations.  The goal
of the present Letter is to explore whether this novel transition really
exists and whether its features are in agreement with the theoretical
predictions.

Since one component systems are very prone to crystallization we have
simulated a binary system of dumbbells. For the sake of computational
efficiency these dumbbells are not hard core particles but instead
interact with a Lennard-Jones (LJ) potential. Thus the system we consider
is a binary mixture of rigid, symmetric dumbbell molecules, to be denoted
by $AA$ and $BB$ dumbbells. Each molecule consists of two identical fused
LJ particles of type $A$ or $B$ having the same mass $m$, and their bond
lengths are denoted by $l_{AA}$ and $l_{BB}$. The interaction between two
molecules is given by the sum of the LJ interactions between the four
constituent sites, $V_{\alpha \beta}(r) = 4 \epsilon_{\alpha \beta}
\{ (\sigma_{\alpha \beta}/r)^{12} - (\sigma_{\alpha \beta}/r)^{6}
\} + C_{\alpha \beta} + D_{\alpha \beta} r$,
where $\alpha, \beta \in \{A,B\}$, with the LJ parameters 
$\epsilon_{AA} = 1.0$, $\sigma_{AA} = 1.0$, $\epsilon_{AB} = 1.5$,
$\sigma_{AB} = 0.8$, $\epsilon_{BB} = 0.5$, and $\sigma_{BB} = 0.88$,
taken from Ref.~\cite{Kob94}. $C_{\alpha \beta}$ and 
$D_{\alpha \beta}$ are chosen so that $V_{\alpha \beta}(r)$ and 
$V_{\alpha \beta}^{\prime}(r)$ are zero at the cutoff
$r_{\rm cut} = 2.5 \, \sigma_{\alpha \beta}$,
and the LJ parameters are simultaneously renormalized 
to ensure that the minimum of the potential is
$-\epsilon_{\alpha \beta}$ at $r = 2^{1/6} \, \sigma_{\alpha \beta}$.
Bond lengths are specified
by a single elongation parameter $\zeta$, defined such that $\zeta =
l_{AA} / \sigma_{AA} = l_{BB} / \sigma_{BB}$.  The number of $AA$
and $BB$ dumbbells is 800 and 200, respectively, giving the number
densities $\rho_{AA} = 800 / L^{3}$ and $\rho_{BB} = 200 / L^{3}$,
where $L$ denotes the length of the cubic box used in the simulation.
The ``packing fraction'' is then defined by $\varphi_{\rm tot} =
\varphi_{AA} + \varphi_{BB}$, where $\varphi_{\alpha\alpha} = (\pi/6)
\rho_{\alpha\alpha} \sigma_{\alpha\alpha}^{3} (1+\frac{3}{2} \zeta -
\frac{1}{2} \zeta^{3})$ for $0 \le \zeta \le 1$.  
Fixing $\varphi_{\rm tot}$ specifies thus $L$
for each $\zeta$. In this study we have used the value $\varphi_{\rm
tot} = 0.708$. We note that, when specialized to $\zeta = 0$, our system
reduces to the spherical system studied in Ref.~\cite{Kob94}, apart from a
factor of 2 in temperature $T$ and mass.  In the following, all quantities
are expressed in reduced units with the unit of length $\sigma_{AA}$, the
unit of energy $\epsilon_{AA}$ (setting $k_{B} = 1$), and the unit of time
$(m \sigma_{AA}^{2} / \epsilon_{AA})^{1/2}$. Standard molecular-dynamics
simulations have been performed like in Ref.~\cite{Kob94}.  The longest
runs were 10$^{8}$ time steps and in order to improve the statistics
of the results we have averaged over 16 independent runs.  More details
will be presented elsewhere~\cite{Chong-Kob-2}.

\begin{figure}[tb]
\includegraphics[width=0.7\linewidth]{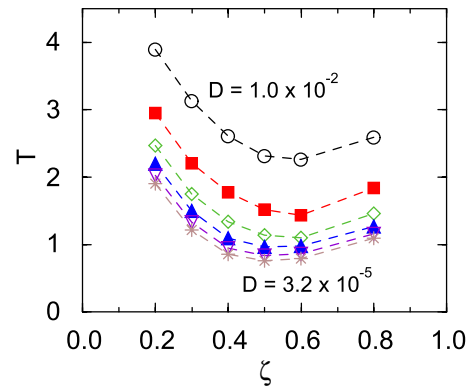}
\caption{
Symbols, connected by lines, show the center of mass 
iso-diffusivity curves for the $AA$
dumbbells. The diffusion constants $D$ are
$D = 1.0 \times 10^{-2}$,
$3.2 \times 10^{-3}$,
$1.0 \times 10^{-3}$,
$3.2 \times 10^{-4}$,
$1.0 \times 10^{-4}$, and
$3.2 \times 10^{-5}$
from top to bottom.}
\label{fig:1}
\end{figure}

It is necessary to make a comment on the notion of ``packing fraction''.
Since LJ interactions are finite, the LJ dumbbells can penetrate each
other making thus the concept of packing fraction somewhat ill-defined.
However, it is reasonable to assume that, at fixed $T$, this penetration
is only a very weak function of the elongation $\zeta$. Therefore it does
make sense to study the relaxation dynamics of the system at given $T$
and $\varphi_{\rm tot}$ as a function of $\zeta$ and hence to check for
the presence of a non-monotonous $\zeta$-dependence of the diffusion
constant and the predicted type-A transition.

To characterize the $\zeta$-dependence of the translational dynamics
we have calculated the mean-squared displacement of the center
of mass of the molecules, from which the diffusion constant $D$
is determined.  In Fig.~\ref{fig:1} we show iso-diffusivity loci, at
fixed $\varphi_{\rm tot}$, as a function of $\zeta$.  Since, for $D \to
0$, the iso-diffusivity lines approach the glass transition locus, they
provide a numerical estimate of the shape and location of the glass line,
which can be compared with theoretical predictions.

From this figure we recognize that at fixed $T$ the diffusion
constant increases with increasing $\zeta$, reaches a maximum at
around $\zeta=0.5$ and then decreases again. Since this behavior is
observed for all values of $D$ we can conclude that the temperature
of dynamic arrest does indeed show a minimum as a function of $\zeta$,
in agreement with the experimental findings~\cite{Donev04,Williams03}
and the theoretical calculations for a system of hard dumbbells
(HDS)~\cite{Chong-HDS}.  In Ref.~\cite{Chong-HDS} it has been argued that
such a $\zeta$-dependence can be understood from the $\zeta$-dependence
of the first sharp diffraction peak in the static structure factor and in
Ref.~\cite{Chong-Kob-2} we will demonstrate that the proposed mechanism
seems indeed relevant for the present system.

\begin{figure}[tb]
\includegraphics[width=0.6\linewidth]{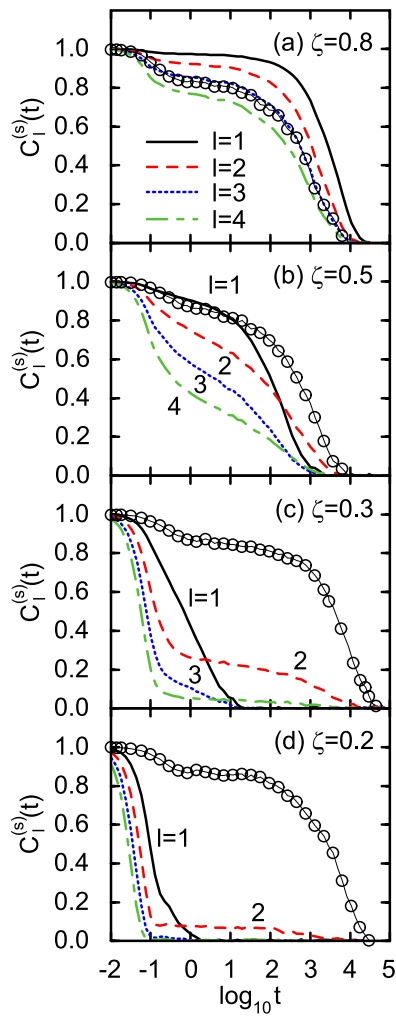}
\caption{
The coherent intermediate scattering function (circles)
and the reorientational correlators $C_{\ell}^{\rm (s)}(t)$ 
for the $AA$ dumbbells in supercooled states 
for $\ell = 1$ (solid line), 2 (dashed
line), 3 (dotted line), and 4 (dash-dotted line) for
(a) $\zeta = 0.8$ and $T = 1.06$,
(b) $\zeta = 0.5$ and $T = 0.75$, 
(c) $\zeta = 0.3$ and $T = 1.13$, and
(d) $\zeta = 0.2$ and $T = 1.85$.}
\label{fig:2}
\end{figure}

We now turn our attention to the $\zeta$-dependence of the rotational
dynamics that we will characterize in terms of the above defined
reorientational correlators $C_{\ell}^{\rm (s)}(t)$.  The time dependence
of these correlators is shown in Fig.~\ref{fig:2} for the case $\ell =
1,...,4$.  Different panels correspond to four representative values of
$\zeta$, and the temperatures are close to those given by the $D = 3.2
\times 10^{-5}$ iso-diffusivity locus.  At this $D$ value, the lowest
curve reported in Fig.~\ref{fig:1}, the system is strongly 
supercooled~\cite{supercooled}.
For the sake of comparison we have also included, as circles, the
normalized coherent intermediate scattering function $F(q,t)$ for the
wave-vector $q$ that corresponds to the first maximum in the static
structure factor~\cite{Chong-Kob-2}.

Most of these correlators show a two-step relaxation: They decay on
intermediate times to a plateau, a motion that corresponds to the
relaxation of the molecule inside the cage formed by its neighbors,
followed by a final $\alpha$-decay in which the molecules leave their
cage (or reorient in the case of $C_{\ell}^{\rm (s)}(t)$).  In the
following we will denote the plateau height by $f_{\ell}^{\rm c}$ and
estimate its value via appropriate fits of $C_l^{\rm (s)}(t)$ with von
Schweidler's law~\cite{Goetze91b,Chong-Kob-2}.  The $\alpha$-relaxation
time $\tau_{\ell}$ will be defined as $C_{\ell}^{\rm (s)}(\tau_{\ell})
= f_{\ell}^{\rm c}/10$.

For sufficiently large elongation ($\zeta = 0.8$, Fig.~\ref{fig:2}a),
steric hindrance for reorientations is substantial, and thus
$C_{\ell}^{\rm (s)}(t)$ exhibit the dynamics of the strong steric
hindrance scenario~\cite{Chong-HDS,Goetze00c}: The plateau height
$f_{\ell}^{\rm c}$ and the $\alpha$-relaxation time $\tau_{\ell}$ decrease
monotonously with increasing $\ell$.  In particular, $f_{1}^{\rm c}
> f_{2}^{\rm c}$ and $\tau_{1} > \tau_{2}$.  These features are in
qualitative agreement with previous findings for various glass-forming
systems~\cite{PC-strong-all,glycerol-strong-all, Fabbian98, Rinaldi01}.
Upon a decrease of the elongation, the steric hindrance effects
for reorientations weaken, leading to smaller $f_{\ell}^{\rm c}$
compared to those for larger elongation. The relations $f_{1}^{\rm c} >
f_{2}^{\rm c}$ and $\tau_{1} > \tau_{2}$ hold until $\zeta \approx 0.6$.
For intermediate elongations, e.g. $\zeta=0.5$ shown in Fig.~\ref{fig:2}b,
the condition $f_{1}^{\rm c} > f_{2}^{\rm c}$ still holds, but now
$\tau_{1} < \tau_{2}$.  Thus for intermediate elongations one expects
that the strength of the $\alpha$-peak in dielectric spectroscopy -- a
technique sensitive to the $\ell=1$ decorrelation -- is more pronounced
than the one in dynamic light scattering ($\ell = 2$), but that the
location of the former is at higher frequencies than the latter.

If the elongation is decreased even further, Figs.~\ref{fig:2}c and
\ref{fig:2}d, the $\alpha$-relaxation process for $\ell=1$ basically
vanishes, although the one for $\ell=2$ is still clearly visible.
In this case  $f_{1}^{\rm c} = 0$.  A closer inspection of the curves
shows that this result holds for all odd values of $\ell$.  Thus, for
small $\zeta$, a dielectric spectroscopy experiment would not show anymore
an $\alpha$-peak, although such a peak would still be visible in a dynamic
light scattering experiment or an inelastic neutron scattering experiment
(which probes $F(q,t)$).  Results reported in Fig.~\ref{fig:2} provide
evidence that, for nearly the same value of $D$, the relaxation dynamics
of the molecules depends strongly on their shape: For elongated molecules
all correlators decay on a similar time scale, whereas for more spherical
molecules the translational and even-$\ell$ reorientational dynamics
relax much slower than the reorientational ones for odd values of $\ell$.

The observed peculiar behavior of the odd-$\ell$ reorientational
correlators at small values of $\zeta$ can be interpreted as being
caused by the nearby type-A transition, predicted by MCT for hard
dumbbells~\cite{Chong-HDS}.  Evidence for the existence of this
transition can be obtained by investigating the $\zeta$-dependence of
the height of the plateaus in the orientational correlation functions,
i.e. $f_{\ell}^{\rm c}(\zeta)$. MCT predicts that, moving along the
ideal glass line, for even $\ell$ this height decreases smoothly with
decreasing $\zeta$ and becomes zero at $\zeta=0$, reflecting the weakened
steric hindrance effects.  In contrast to this, the $\zeta$-dependence
of $f_{\ell}^{\rm c}(\zeta)$ for odd $\ell$ is predicted to show a
smooth dependence on $\zeta$ down to the critical value $\zeta_{\rm c}$
discussed above, to vanish at $\zeta=\zeta_{\rm c}$ and then to stay
zero for $\zeta < \zeta_{\rm c}$.

\begin{figure}[tb]
\includegraphics[width=0.7\linewidth]{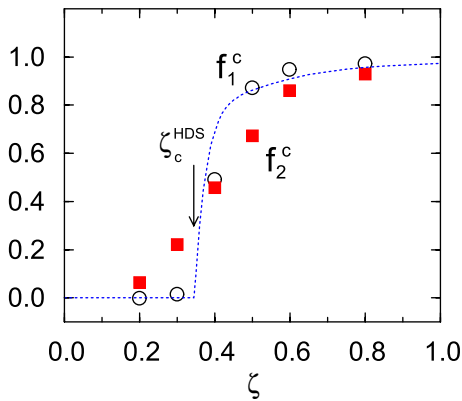}
\caption{
Plateau heights $f_{1}^{\rm c}$ (circles) and $f_{2}^{\rm c}$ 
(filled squares)
for $AA$-dumbbell's correlators $C_{\ell}^{\rm (s)}(t)$ for $\ell = 1$ and
2, respectively. The dotted line represents $f_{1}^{\rm c}$ for the HDS taken
from Ref.~\protect\cite{Chong-HDS} with the arrow indicating the critical
elongation $\zeta_{\rm c}^{\rm HDS}$.}
\label{fig:3}
\end{figure}

These predictions are indeed compatible with the results from our
simulations as demonstrated in Fig.~\ref{fig:3} where we plot the
$\zeta$-dependence of $f_{1}^{\rm c}$ and $f_{2}^{\rm c}$.  Starting from
large $\zeta$, the plateau height $f_{2}^{\rm c}$ decreases smoothly
toward the limit $f_{2}^{\rm c} = 0$ for spherical system ($\zeta
= 0$).  Instead, the height $f_{1}^{\rm c}$ decreases smoothly with
decreasing $\zeta$ at large and intermediate $\zeta$, but it shows,
around $\zeta=0.4$, a sudden drop and becomes zero within the statistical
accuracy around $\zeta = 0.3$.  This behavior suggests the existence of a
critical elongation $\zeta_{\rm c}$ which is located between $\zeta = 0.3$
and 0.4.  For the sake of comparison we have added to Fig.~\ref{fig:3}
also the theoretical curve $f_{1}^{\rm c}$ for the HDS (dotted line)
as well as the location of its critical elongation $\zeta_{\rm c}^{\rm
HDS}$~\cite{Chong-HDS}.  Note that $f_{1}^{\rm c}(\zeta)$ shows nearly
the same $\zeta$-dependence as the theoretical quantity for HDS, which
corroborates the existence of $\zeta_{\rm c}$ between $\zeta = 0.3$
and 0.4, and hence the presence of the type-A transition in molecular
systems of small elongations.

Since the appearance of the plateau regime for the even-$\ell$ correlators
is due to the cage effect, the results from the simulation for $\zeta =
0.2$ and 0.3 (Figs.~\ref{fig:2}c and \ref{fig:2}d) imply that (i) the
relaxation of the odd-$\ell$ correlators occurs via large-angle flips
between the energetically identical orientations ${\vec e}$ and $-{\vec
e}$, which do not alter the even-$\ell$ correlators, and (ii) these flips
occur before the molecule has left the cage. Such a behavior for the
reorientational dynamics is in striking contrast to the strong steric
hindrance scenario for large elongations for which the reorientation
occurs continuously and on the time scale of the molecule's escape from
the cage.

Even for elongations exceeding $\zeta_{\rm c}$, precursor effects
of the type-A transition do influence seriously the results for the
odd-$\ell$ reorientational dynamics, the most noticeable one being the
speeding up of their $\alpha$-processes~\cite{Goetze00c}. For instance,
one concludes from Fig.~\ref{fig:2}b that the steric hindrance effect
is still considerable for $\zeta = 0.5$ since $f_{1}^{\rm c}$ remains
close to one. However, the canonical order $\tau_{1} > \tau_{2}$ expected
for the strong steric hindrance scenario is violated for $\zeta = 0.5$,
as can be seen from Fig.~\ref{fig:2}b, which is evidence for the nearby
type-A transition.

Although strictly speaking the type-A transition discussed here exists
only for symmetric molecules in which two ends are identical,
most of the mentioned features
can be expected to be observable, in the form of precursor effects, also
for molecules in which this symmetry is slightly broken. E.g. K\"ammerer
{\em et al.} have studied a system with slightly {\em asymmetric}
dumbbells with $\zeta = 0.5$~\cite{Kaemmerer97}. These authors found
$\tau_{1} < \tau_{2}$ in spite of the large plateau $f_{1}^{\rm c}$, in
full agreement with the present results and the theoretical expectation,
indicating that the scenario caused by a nearby type-A transition
can indeed be found in a class of molecular systems which are
weakly elongated.

We thank W. G\"otze for comments on the
manuscript. Part of this work has been supported by the European
Community's Human Potential Programme under contract HPRN-CT-2002-00307,
DYGLAGEMEM.


\begin{thebibliography}{99}
\bibitem{comment-zeta-alpha}
$\zeta$ is related to the
aspect ratio \protect$\alpha$ via \protect$\alpha = 1 + \zeta$.
\bibitem{supercooled}
Since the thermodynamic melting points have not been
investigated for our model, we use the word ``supercooled''
to indicate that the decay of time-correlation functions
shows a two-step relaxation. This evidence of slow pre-glassy
dynamics is often used to generalize in a dynamic sense the
concept of supercooled states.
\end{thebibliography}
\end{document}